\documentclass[aps,prd,onecolumn,groupedaddress,showpacs,nofootinbib,amssymb]{revtex4}
\usepackage[dvips]{graphicx}
\usepackage{amssymb}
\usepackage{amsmath}
\usepackage{graphicx,,color}
\usepackage{amsfonts}
\usepackage{bm}
\usepackage{cancel}
\usepackage{comment}

\newcommand\be{\begin{equation}}
\newcommand\ee{\end{equation}}

\allowdisplaybreaks[4]

\begin{document}

\tolerance=5000

\title{Non-minimally Coupled Scalar $k$-Inflation Dynamics}
\author{V.K.~Oikonomou,$^{1,2,3,*}$\,}

\affiliation{$^{1)}$ Department of Physics, Aristotle University
of Thessaloniki, Thessaloniki 54124,
Greece\\
$^{2)}$ Laboratory for Theoretical Cosmology, Tomsk State
University of Control Systems and Radioelectronics, 634050 Tomsk,
Russia (TUSUR)\\
$^{3)}$ Tomsk State Pedagogical University, 634061 Tomsk, Russia\\
$^{*}$ Corresponding author: voikonomou@auth.gr}

\tolerance=5000

\begin{abstract}
In this work we shall study $k$-inflation theories with
non-minimal coupling of the scalar field to gravity, in the
presence of only a higher order kinetic term of the form $\sim
\mathrm{const}\times X^{\mu}$, with
$X=\frac{1}{2}\partial_{\mu}\phi\partial^{\mu}\phi$. The study
will be focused in the cases where a scalar potential is included
or is absent, and the evolution of the scalar field will be
assumed to satisfy the slow-roll or the constant-roll condition.
In the case of the slow-roll models with scalar potential, we
shall calculate the slow-roll indices, and the corresponding
observational indices of the theory, and we demonstrate that the
resulting theory is compatible with the latest Planck data. The
same results are obtained in the constant-roll case, at least in
the presence of a scalar potential. In the case that models
without potential are considered, the results are less appealing
since these are strongly model dependent, and at least for a
power-law choice of the non-minimal coupling, the theory is
non-viable. Finally, due to the fact that the scalar and tensor
power spectra are conformal invariant quantities, we argue that
the Einstein frame counterpart of the non-minimal $k$-inflation
models with scalar potential, can be a viable theory, due to the
conformal invariance of the observational indices. The Einstein
frame theory is more involved and thus more difficult to work with
it analytically, so one implication of our work is that we provide
evidence for the viability of another class of $k$-inflation
models.
\end{abstract}


\maketitle

\section{Introduction}

The striking GW170817 event \cite{GBM:2017lvd} in 2017 verified in
a direct way the propagation of astrophysical gravitational waves
emitted from the merging of compact astrophysical objects. Apart
from generating a new scientific direction in modern astrophysics,
that of gravitational wave astronomy, the GW170817 event had
significant theoretical implications in the field of astrophysics
and cosmology, due to the fact that it imposed serious
restrictions on several modified gravity theoretical frameworks.
Particularly, the most striking result from a theoretical point of
view was the fact that the gravitational waves and the gamma rays
emitted from the merging of the two neutron stars arrived
simultaneously, so this result ruled out many theories of
Horndeski type, see Ref. \cite{Ezquiaga:2017ekz} for an extensive
analysis along this research line.

However, there are still many modified gravity theories which
still yield a gravitational wave speed $c_T^2=1$ in natural units,
such as $f(R)$ gravity, $f(G)$ Gauss-Bonnet theories and so on.
For an extensive account on modified gravity theories, we refer
the reader to Refs.
\cite{Nojiri:2017ncd,Nojiri:2010wj,Nojiri:2006ri,Capozziello:2011et,Capozziello:2010zz,delaCruzDombriz:2012xy,Olmo:2011uz}.
In addition, modified gravity theories can actually provide a
theoretical framework that makes possible the description and in
some cases the unification of the two acceleration eras of our
Universe, such as $f(R)$ gravity for example. The modification of
Einstein-Hilbert gravity seems compelling, at least when
cosmological scales are considered. In fact, although
Einstein-Hilbert gravity seems to describe accurately to a great
extent the current perception of astrophysical scales physics, it
seems that at large scales needs to be modified. Indeed, the dark
energy era in the context of simple Einstein-Hilbert gravity
requires the introduction of phantom scalar fields, and exotic
phenomena follow, such us the existence of late-time finite time
singularities. In the context of modified gravity, the
inflationary era and the late-time acceleration era can be
harbored under the unified framework of a single theory
\cite{Nojiri:2003ft}.

One class of the remaining viable gravitational theories after the
GW170817 event, is the $k$-inflation class of theories, see Refs.
\cite{ArmendarizPicon:1999rj,Chiba:1999ka,ArmendarizPicon:2000dh,Matsumoto:2010uv,ArmendarizPicon:2000ah,Chiba:2002mw,Malquarti:2003nn,Malquarti:2003hn,Chimento:2003zf,Chimento:2003ta,Scherrer:2004au,Aguirregabiria:2004te,ArmendarizPicon:2005nz,Abramo:2005be,Rendall:2005fv,Bruneton:2006gf,dePutter:2007ny,Babichev:2007dw,Deffayet:2011gz,Kan:2018odq,Unnikrishnan:2012zu,Li:2012vta}
for an important stream of research articles. Also it is notable
that $k$-inflation models can be related to certain classes of
Palatini $f(R)$ gravity, see for example \cite{Gialamas:2019nly}.
These $k$-inflation theories can actually describe both the
inflationary era, and also the late-time acceleration era, and are
also known as $k$-essence theories. In this paper we shall focus
on the inflationary phenomenology of a modified version of
$k$-inflation theories, for which we shall assume that a
non-minimal coupling between the scalar curvature and the scalar
field exists, at least in most of the cases. We shall also examine
the above theories in the presence of a scalar potential and in
the absence of a scalar potential, and more importantly, we shall
assume that the standard kinetic term of the scalar field is
absent, and only higher powers of
$X=\frac{1}{2}\partial_{\mu}\phi\partial^{\mu}\phi$ are present in
the Lagrangian. We shall perform an extensive phenomenological
analysis of the various models, and we shall confront each of them
with the observational data coming from the 2018 Planck
collaboration \cite{Akrami:2018odb}. For the models we shall
study, either the slow-roll or the constant-roll condition will be
assumed. The slow-roll condition is needed in order to provide a
sufficiently long inflationary era in order to solve the standard
inflation phenomenological problems
\cite{Guth:1980zm,Starobinsky:1982ee,Linde:1983gd,Albrecht:1982wi},
while the constant-roll is a relatively new phenomenological
condition in inflationary cosmology, and it is appealing since it
may generate non-Gaussianities in the power spectrum of even
scalar theories of gravity, see Refs.
\cite{Inoue:2001zt,Tsamis:2003px,Kinney:2005vj,Tzirakis:2007bf,
Namjoo:2012aa,Martin:2012pe,Motohashi:2014ppa,Cai:2016ngx,
Motohashi:2017aob,Hirano:2016gmv,Anguelova:2015dgt,Cook:2015hma,
Kumar:2015mfa,Odintsov:2017yud,Odintsov:2017qpp,Lin:2015fqa,Gao:2017uja,Nojiri:2017qvx,Oikonomou:2017bjx,Odintsov:2017hbk,Oikonomou:2017xik,Cicciarella:2017nls,Awad:2017ign,Anguelova:2017djf,Ito:2017bnn,Karam:2017rpw,Yi:2017mxs,Mohammadi:2018oku,Gao:2018tdb,Mohammadi:2018wfk,Morse:2018kda,Cruces:2018cvq,GalvezGhersi:2018haa,Boisseau:2018rgy,Gao:2019sbz,Lin:2019fcz,Mohammadi:2019qeu}
for an important stream of papers in the subject.

The results of our analysis can be deemed interesting since we
found that the non-minimally coupled $k$-inflation theories
without potential can be viable for both the constant and the
slow-roll conditions holding true, and for various forms of the
potential and the non-minimal coupling function chosen. Also in
the potential-less case, we demonstrate that the power-law
non-minimal couplings are phenomenologically excluded, at least
when quadratic order kinetic terms are considered in the
Lagrangian. Finally, we demonstrate that the non-minimally coupled
theories with potential are conformally transformed to
$k$-inflation theories with the presence of a canonical scalar
kinetic term, and also with potential. Due to the fact that the
power spectrum of the scalar and of the tensor perturbations is
invariant under a conformal transformation, we immediately have a
direct connection of the two theories, thus the conformally
transformed theory can also, in principle, be viable.

\section{Non-minimal $k$-inflation Models Phenomenology}

The $k$-inflation theories belong to the general class of theories
of the form $f(R,\phi,X)$ with $X$ being
$X=\frac{1}{2}\partial^{\mu}\phi\partial_{\mu}\phi$. The general
analysis of cosmological perturbations for this kind of theories
was performed in a series of papers which can be found in
Refs.~\cite{Noh:2001ia,Hwang:2005hb,Hwang:2002fp,Kaiser:2013sna},
the notation of which we shall adopt in this paper. The
cosmological geometric background shall be assumed to be a flat
Friedmann-Robertson-Walker (FRW) with line element,
\begin{equation}
\label{metricfrw} ds^2 = - dt^2 + a(t)^2 \sum_{i=1,2,3}
\left(dx^i\right)^2\, ,
\end{equation}
where $a(t)$ denotes the scale factor. The gravitational action
for the general $f(R,\phi,X)$ theory is,
\begin{equation}
\label{mainactionB} \mathcal{S}=\int d^4x\sqrt{-g}\left[
\frac{1}{2}f(R,\phi,X) \right]\, ,
\end{equation}
and the exact form of the function $f(R,X,\phi)$ will be defined
later for the various models which we shall study. However, the
general form of the $f(R,X,\phi)$ function will be the following,
\begin{equation}\label{frxfunction}
f(R,\phi,X)=\frac{h(\phi)R}{\kappa^2}-2V(\phi)+g(\phi) X^{\mu}\, ,
\end{equation}
where in Eq. (\ref{frxfunction}),
$X=\frac{1}{2}\partial^{\mu}\phi\partial_{\mu}\phi$, and also
$V(\phi)$ is the scalar potential, which in some cases will be
assumed to be equal to zero. For simplicity we shall assume that
$\mu=2$, but the arguments that will apply for this subcase,
easily apply for the general case. In addition
$\kappa^2=\frac{1}{M_p^2}$, where $M_p$ is the reduced Planck
mass. By assuming a FRW background, the gravitational equations
for a general form of the function $f(R,X,\phi)$ are,
\begin{equation}\label{euqationsofmotion1}
3H^2F=f_{,X}X+\frac{RF-f}{2}-3H\dot{F}\, ,
\end{equation}
\begin{equation}\label{euqationsofmotion2}
-2\dot{H}F=Xf_{,X}+\ddot{F}-H\dot{F}\, ,
\end{equation}
\begin{equation}\label{euqationsofmotion3}
\frac{1}{a^3}\left( a^3\dot{\phi}f_{,X}\right)^{.}+f_{,\phi}=0\, ,
\end{equation}
where the ``dot'' in the above equations indicates as usual the
differentiation with respect to the cosmic time, and in addition
$F=\frac{\partial f}{\partial R}$. Also, where used in the
following, the prime denotes differentiation with respect to the
assumed argument of the differentiated function.

As we mentioned in the introduction, the cosmological tensor
perturbations propagate with a speed $c_T^2=1$, however the wave
speed $c_A$ of the scalar perturbations have a non-trivial form,
\begin{equation}\label{wavespeed}
c_A^2=\frac{Xf_{,X}+\frac{3\dot{F}^2}{2F}}{Xf_{,X}+2X^2f_{,XX}+\frac{3\dot{F}^2}{2F}}\,
,
\end{equation}
where $f_{,X}=\frac{\partial f}{\partial X}$ and
$f_{,XX}=\frac{\partial^2 f}{\partial X^2}$. The dynamics of
inflation is quantified in terms of the following ``slow-roll''
parameters \cite{Hwang:2005hb},
\begin{align}\label{slowrollparameters}
& \epsilon_1=\frac{\dot{H}}{H^2}\, , \,\,\,
\epsilon_2=\frac{\ddot{\phi}}{H\dot{\phi}}\, ,\,\,\,
\epsilon_3=\frac{\dot{F}}{2HF}\, ,\,\,\,
\epsilon_4=\frac{\dot{E}}{2HE}\, ,
\end{align}
although the terminology ``slow-roll'' for these is not accurate,
since the slow-roll assumption is not imposed, at least for the
moment. We shall impose the slow-roll or the constant-roll
assumption for the models to be presented in the following
sections. Also in the general case, the function $E$ appearing in
the slow-roll index $\epsilon_4$ is defined in general as follows,
\begin{equation}\label{functionE}
E=-\frac{F}{2X}\left(X f_{,X}+2X^2 f_{,XX}+\frac{3\dot{F}^2}{2F}
\right)\, .
\end{equation}
In Ref. \cite{Hwang:2005hb} the tensor and scalar perturbations
for the general $f(R,\phi,X)$ theory were studied in detail, and
the spectral index of the primordial curvature perturbations was
shown that it is possible to express it in terms of the slow-roll
indices, and this reads,
\begin{equation}\label{spectralindex}
n_s=1+2\frac{2\epsilon_1-\epsilon_2+\epsilon_3-\epsilon_4}{1+\epsilon_1}\,
,
\end{equation}
while the tensor-to-scalar ratio is equal to,
\begin{equation}\label{tensortoscalarration}
r=16 |\epsilon_1-\epsilon_3|c_A\, ,
\end{equation}
where the wave speed $c_A$ is defined in Eq. (\ref{wavespeed}) for
the $f(R,X,\phi)$ theory. In the following subsections we shall
focus on confronting the inflationary phenomenology of various
different models of the form (\ref{frxfunction}) with the
observational data, and for each case we shall examine the
possibility of having instabilities in the resulting theory, by
examining the values of the wave speed $c_A^2$, for the values of
the free parameters which guarantee the phenomenological viability
of the theory in each case.

\subsection{Models with Potential: Slow-roll Phenomenology}

The most phenomenologically interesting models from the models
belonging in the class of Eq. (\ref{frxfunction}), are those with
scalar potential, and in this case the functional form of the
$f(R,X,\phi)$ gravity will be,
\begin{equation}\label{frxfunctionpotentialslowroll}
f(R,\phi,X)=\frac{h(\phi)R}{\kappa^2}-2V(\phi)+c_1 X^{2}\, .
\end{equation}
We shall study the case $\mu=2$ and $g(\phi)=c_1$ of Eq.
(\ref{frxfunction}), where $c_1$ is a free parameter in the
theory. The case for general $\mu$ in principle generates the same
phenomenology, but perplexes the calculations to some extent, so
for the sake of simplicity we focus on the case $\mu=2$. For the
function $f(R,X,\phi)$ chosen as in Eq.
(\ref{frxfunctionpotentialslowroll}), we have
$F=\frac{h(\phi)}{\kappa^2}$, and with the slow-roll assumption
made for the scalar field,
\begin{equation}\label{slowrollscalarfield}
\ddot{\phi}\ll H\dot{\phi}\, ,
\end{equation}
and,
\begin{equation}\label{slowrollassumption2}
c_1\dot{\phi}^4\ll V(\phi)\, ,
\end{equation}
the gravitational equations read,
\begin{equation}\label{euqationsofmotion11potentialslowroll}
H^2\simeq \frac{\kappa^2V(\phi)}{3h(\phi)}\, ,
\end{equation}
\begin{equation}\label{euqationsofmotion21potentialslowroll}
\dot{H}\simeq \frac{\kappa^2}{2h(\phi )}\Big{(}\frac{H
h'(\phi)\dot{\phi}}{\kappa^2}-\frac{h''(\phi)\dot{\phi}^2}{\kappa^2}-\frac{c_1\dot{\phi}^4}{2}\Big{)}\,
,
\end{equation}
\begin{equation}\label{euqationsofmotion31potentialslowroll}
-3c_1H\dot{\phi}^3\simeq
2V'(\phi)-\frac{h'(\phi)12H^2}{\kappa^2}\, ,
\end{equation}
where the ``prime'' indicates differentiation with respect to the
scalar field. At this point, since we have assumed a slow-roll
evolution for the scalar field, we shall assume that only the
third term is dominant in Eq.
(\ref{euqationsofmotion21potentialslowroll}), but this must be
explicitly verified in the end of the calculation. Actually, we
also examined the cases that the other two terms dominate the
evolution, but the approximation was not valid, so from now on we
assume that,
\begin{equation}\label{neweqn}
\Big{|}\frac{h''(\phi)\dot{\phi}^2}{\kappa^2} \Big{|}\ll
\Big{|}\frac{c_1\dot{\phi}^4}{2}\Big{|}\, ,
\end{equation}
and,
\begin{equation}\label{neweqn1}
\Big{|}\frac{H h'(\phi)\dot{\phi}}{\kappa^2} \Big{|}\ll
\Big{|}\frac{c_1\dot{\phi}^4}{2}\Big{|} \, ,
\end{equation}
hence, $\dot{H}$ is approximated as,
\begin{equation}\label{pereasei}
\dot{H}\simeq -\frac{c_1\kappa^2\dot{\phi}^4}{2h(\phi)}\, .
\end{equation}
So upon combining the above equations we obtain the slow-roll
expressions of $H$, $\dot{H}$ and $\dot{\phi}$ as a function of
the scalar field, which are,
\begin{equation}\label{euqationsofmotion11potentialslowroll1}
H\simeq \frac{\sqrt{\frac{\kappa ^2 V(\phi )}{h(\phi
)}}}{\sqrt{3}}\, ,
\end{equation}
\begin{equation}\label{euqationsofmotion21potentialslowroll2}
\dot{H}\simeq -\frac{\text{c1} \kappa ^2 \left(\frac{2 V'(\phi
)-\frac{4 V(\phi ) h'(\phi )}{h(\phi )}}{\text{c1}
\sqrt{\frac{\kappa ^2 V(\phi )}{h(\phi )}}}\right)^{4/3}}{4\
3^{2/3} h(\phi )}\, ,
\end{equation}
\begin{equation}\label{euqationsofmotion31potentialslowroll3}
\dot{\phi}\simeq \frac{\sqrt[3]{-\frac{2 V'(\phi )-\frac{4 V(\phi
) h'(\phi )}{h(\phi )}}{\text{c1} \sqrt{\frac{\kappa ^2 V(\phi
)}{h(\phi )}}}}}{\sqrt[6]{3}}\, .
\end{equation}
The slow-roll indices (\ref{slowrollparameters}) as functions of
the scalar field $\phi$, can easily be evaluated by using
equations (\ref{euqationsofmotion11potentialslowroll1}),
(\ref{euqationsofmotion21potentialslowroll2}) and
(\ref{euqationsofmotion31potentialslowroll3}), and these read,
\begin{align}\label{slowrollparameters}
& \epsilon_1\simeq -\frac{\sqrt[3]{3} \text{c1} \left(\frac{2
V'(\phi )-\frac{4 V(\phi ) h'(\phi )}{h(\phi )}}{\text{c1}
\sqrt{\frac{\kappa ^2 V(\phi )}{h(\phi )}}}\right)^{4/3}}{4 V(\phi
)}\, , \\ \notag &
 \epsilon_2\simeq 0\,
,\\ \notag & \epsilon_3\simeq \frac{h'(\phi ) \sqrt[3]{\frac{6
V(\phi ) h'(\phi )-3 h(\phi ) V'(\phi )}{\text{c1} h(\phi )
\sqrt{\frac{\kappa ^2 V(\phi )}{h(\phi )}}}}}{2^{2/3} h(\phi )
\sqrt{\frac{\kappa ^2 V(\phi )}{h(\phi )}}}\, ,\\ \notag &
\epsilon_4\simeq \frac{\sqrt[3]{6} h'(\phi ) \left(\frac{2 V(\phi
) h'(\phi )-h(\phi ) V'(\phi )}{\text{c1} h(\phi )
\sqrt{\frac{\kappa ^2 V(\phi )}{h(\phi )}}}\right)^{2/3}
\left(\text{c1} \kappa ^2 \left(\frac{2 V(\phi ) h'(\phi )-h(\phi
) V'(\phi )}{\text{c1} h(\phi ) \sqrt{\frac{\kappa ^2 V(\phi
)}{h(\phi )}}}\right)^{2/3}+\sqrt[3]{6} h''(\phi
)\right)}{\sqrt[3]{6} h'(\phi )^2 \sqrt{\frac{\kappa ^2 V(\phi
)}{h(\phi )}} \sqrt[3]{\frac{2 V(\phi ) h'(\phi )-h(\phi ) V'(\phi
)}{\text{c1} h(\phi ) \sqrt{\frac{\kappa ^2 V(\phi )}{h(\phi
)}}}}+4 \kappa ^2 V(\phi ) h'(\phi )-2 \kappa ^2 h(\phi ) V'(\phi
)}\, .
\end{align}
In order to obtain the spectral index of the primordial scalar
curvature perturbations, we need to express the scalar field as a
function of the $e$-foldings number. The slow-roll indices must be
evaluated at the value of the scalar field when the first horizon
crossing occurs during the inflationary era, for which $k=Ha$ and
$\phi=\phi_k$. Also the end of the inflationary era is going to be
determined when $\epsilon_1(\phi_f)=\mathcal{O}(1)$. The
$e$-foldings number is defined as,
\begin{equation}\label{efoldings1cosmictime}
N=\int_{t_k}^{t_f}H(t)dt\, ,
\end{equation}
with $t_k$ is the horizon crossing time instance, and $t_f$ is the
time instance where the inflationary era ends. We can express the
$e$-foldings integral as a function of the scalar field, and we
have,
\begin{equation}\label{efoldings2}
N=\int_{\phi_k}^{\phi_f}\frac{H}{\dot{\phi}}d\phi\, ,
\end{equation}
so by using Eqs. (\ref{euqationsofmotion11potentialslowroll1}),
(\ref{euqationsofmotion21potentialslowroll2}) and
(\ref{euqationsofmotion31potentialslowroll3}), we have,
\begin{equation}\label{efoldings3}
N=\int_{\phi_k}^{\phi_f}\left( \frac{\sqrt{\frac{\kappa ^2 V(\phi
)}{h(\phi )}}}{\sqrt[3]{3} \sqrt[3]{\frac{\frac{4 V(\phi ) h'(\phi
)}{h(\phi )}-2 V'(\phi )}{\text{c1} \sqrt{\frac{\kappa ^2 V(\phi
)}{h(\phi )}}}}}\right)d\phi\, .
\end{equation}
Upon integrating the above, one can have the value of the scalar
field $\phi_k$ as a function of $N$ and $\phi_f$. Also, by solving
the equation $\epsilon_1(\phi_f)\simeq 1$, we may have the value
$\phi_f$ as a function of the free parameters of the theory, so by
using this, we may express the slow-roll indices as functions of
the $e$-foldings number $N$ and the free parameters of the theory.
Accordingly, the observational indices (\ref{spectralindex}) and
(\ref{tensortoscalarration}) can be obtained in closed form.
Eventually, the theoretical results must be compared with the
latest Planck observational data \cite{Akrami:2018odb}, which
constrain the spectral index and the tensor-to-scalar ratio as
follows,
\begin{equation}\label{observationaldatanewresults}
n_s=0.962514\pm 0.00406408,\,\,\,r<0.064\, .
\end{equation}
What now remains is to find an appropriate model which may yield a
viable phenomenology. We have examined two models, firstly the
exponential type of model, with,
\begin{equation}\label{exponentialmodel}
h(\phi)=\Lambda  \exp (\gamma  \kappa  \phi ),\,\,\,V(\phi)=\xi
\exp (-\kappa  \phi )\, ,
\end{equation}
where $\Lambda$, $\gamma$ and $\xi$ are free parameters. The
second model is,
\begin{equation}\label{powerlawmodel}
h(\phi)=\Lambda  \phi ^n,\,\,\,V(\phi)=\xi  \phi ^m\, ,
\end{equation}
with $\xi$ and $\Lambda$ being free parameters. The exponential
type model does not yield a viable phenomenology, at least in the
slow-roll case, but as we show in a later section, it provides a
viable phenomenology only in the constant-roll case. However, the
power-law type of model (\ref{powerlawmodel}) does yield a viable
phenomenology in the slow-roll case, as we now demonstrate. So in
the rest of this subsection, we focus on the power-law type of
model of Eq. (\ref{powerlawmodel}), so for the model
(\ref{powerlawmodel}), the slow-roll indices of Eq.
(\ref{slowrollparameters}) become,
\begin{align}\label{slowrollparametersviablemodel}
& \epsilon_1\simeq -\frac{\sqrt[3]{3} \Lambda ^{2/3} (m-2 n)^{4/3}
\phi ^{\frac{1}{3} (-m+2 n-4)}}{2^{2/3} \sqrt[3]{\text{c1}} \kappa
^{4/3} \sqrt[3]{\xi }}\, , \\ \notag &
 \epsilon_2\simeq 0\,
,\\ \notag & \epsilon_3\simeq -\frac{\sqrt[3]{3} \Lambda ^{2/3} n
\sqrt[3]{m-2 n} \phi ^{\frac{1}{3} (-m+2 n-4)}}{2^{2/3}
\sqrt[3]{\text{c1}} \kappa ^{4/3} \sqrt[3]{\xi }} \, ,\\ \notag &
\epsilon_4\simeq \frac{n (m-2 n) \left(6^{2/3} \text{c1} \kappa ^2
\phi ^2 \left(\frac{\xi  (m-2 n) \phi ^{m-1}}{\text{c1}
\sqrt{\frac{\kappa ^2 \xi  \phi ^{m-n}}{\Lambda }}}\right)^{2/3}+6
\Lambda  n^2 \phi ^n-6 \Lambda  n \phi ^n\right)}{\sqrt[3]{6}
\text{c1} \kappa ^2 \phi ^2 \sqrt[3]{\frac{\xi  (m-2 n) \phi
^{m-1}}{\text{c1} \sqrt{\frac{\kappa ^2 \xi  \phi ^{m-n}}{\Lambda
}}}} \left(n \left(4 \phi  \sqrt{\frac{\kappa ^2 \xi  \phi
^{m-n}}{\Lambda }}-\sqrt[3]{6} n \sqrt[3]{\frac{\xi  (m-2 n) \phi
^{m-1}}{\text{c1} \sqrt{\frac{\kappa ^2 \xi  \phi ^{m-n}}{\Lambda
}}}}\right)-2 m \phi  \sqrt{\frac{\kappa ^2 \xi  \phi
^{m-n}}{\Lambda }}\right)} \, .
\end{align}
Accordingly, the spectral index of the primordial curvature scalar
perturbations $n_s$ and the tensor-to-scalar ratio as a function
of the scalar field can easily be evaluated, but we do not quote
here their final form for brevity. Also, the wave speed $c_A$
appearing in Eq. (\ref{wavespeed}) reads,
\begin{equation}\label{wavespeedmodel1potential}
c_A=\sqrt{\frac{\frac{3^{2/3} \Lambda  n^2 \phi ^{n-2}
\left(\frac{2 m \xi  \phi ^{m-1}-4 n \xi  \phi ^{m-1}}{\text{c1}
\sqrt{\frac{\kappa ^2 \xi  \phi ^{m-n}}{\Lambda
}}}\right)^{2/3}}{2 \kappa ^2}+\frac{\text{c1} \left(\frac{2 m \xi
\phi ^{m-1}-4 n \xi  \phi ^{m-1}}{\text{c1} \sqrt{\frac{\kappa ^2
\xi  \phi ^{m-n}}{\Lambda }}}\right)^{4/3}}{2\
3^{2/3}}}{\frac{3^{2/3} \Lambda  n^2 \phi ^{n-2} \left(\frac{2 m
\xi  \phi ^{m-1}-4 n \xi  \phi ^{m-1}}{\text{c1}
\sqrt{\frac{\kappa ^2 \xi  \phi ^{m-n}}{\Lambda
}}}\right)^{2/3}}{2 \kappa ^2}+\frac{1}{2} \sqrt[3]{3} \text{c1}
\left(\frac{2 m \xi  \phi ^{m-1}-4 n \xi  \phi ^{m-1}}{\text{c1}
\sqrt{\frac{\kappa ^2 \xi  \phi ^{m-n}}{\Lambda
}}}\right)^{4/3}}}\, .
\end{equation}
Having the above at hand, one needs to evaluate these at the value
of the scalar field when the horizon crossing occurs during
inflation, and express everything as a function of the
$e$-foldings number $N$. To this end, we need to evaluate the
integral of Eq. (\ref{efoldings3}), which for the case at hand is,
\begin{equation}\label{numericalefoldingspro}
N=\int_{\phi_k}^{\phi_f}\frac{\sqrt[3]{\text{c1}} \kappa ^{4/3}
\xi ^{5/6} \phi ^{\frac{1}{3} (m-2 n+1)}}{\sqrt[3]{6} \Lambda
^{2/3} \left(\sqrt[3]{2} \sqrt[3]{n}-\sqrt[3]{m}\right)}d\phi \, ,
\end{equation}
so upon integrating we obtain,
\begin{equation}\label{numericalintegrationofmodel1}
N=\frac{3^{2/3} \sqrt[3]{\text{c1}} \kappa ^{4/3} \xi ^{5/6}
\phi_f^{\frac{1}{3} (m-2 n+4)}}{\sqrt[3]{2} \Lambda ^{2/3}
\left(\sqrt[3]{2} \sqrt[3]{n}-\sqrt[3]{m}\right) (m-2
n+4)}-\frac{3^{2/3} \sqrt[3]{\text{c1}} \kappa ^{4/3} \xi ^{5/6}
\phi_k^{\frac{1}{3} (m-2 n+4)}}{\sqrt[3]{2} \Lambda ^{2/3}
\left(\sqrt[3]{2} \sqrt[3]{n}-\sqrt[3]{m}\right) (m-2 n+4)}\, .
\end{equation}
The value of the scalar field $\phi_f$ at the end of the
inflationary era can easily be evaluated by solving the equation
$\epsilon_1(\phi_f)=1$, and it is,
\begin{equation}\label{valueofphifinalmodel1}
\phi_f=\left(\frac{\sqrt[3]{3} \Lambda ^{2/3} m \sqrt[3]{m-2
n}}{2^{2/3} \sqrt[3]{\text{c1}} \kappa ^{4/3} \sqrt[3]{\xi
}}-\frac{\sqrt[3]{6} \Lambda ^{2/3} n \sqrt[3]{m-2
n}}{\sqrt[3]{\text{c1}} \kappa ^{4/3} \sqrt[3]{\xi
}}\right)^{\frac{12}{5 m-9 n+16}}\, ,
\end{equation}
so upon substituting Eq. (\ref{valueofphifinalmodel1}) in Eq.
(\ref{numericalintegrationofmodel1}) and solving with respect to
the value of the scalar field $\phi_k$, we have the latter
expressed as a function of the $e$-foldings number and as a
function of the free parameters of the theory. Finally, by
substituting $\phi_k$ in the slow-roll indices
(\ref{slowrollparametersviablemodel}) and in the wave speed
(\ref{wavespeedmodel1potential}) we may substitute in the
expressions for the spectral index (\ref{spectralindex}) and the
tensor-to-scalar ratio (\ref{tensortoscalarration}), and evaluate
explicitly their functional form as functions of the free
parameters of the model and as a function of the $e$-foldings
number. Their resulting expressions are too long to quote them
here, but let us proceed confronting the model with the latest
Planck constraints (\ref{observationaldatanewresults}).

\begin{figure}[h!]
\centering
\includegraphics[width=18pc]{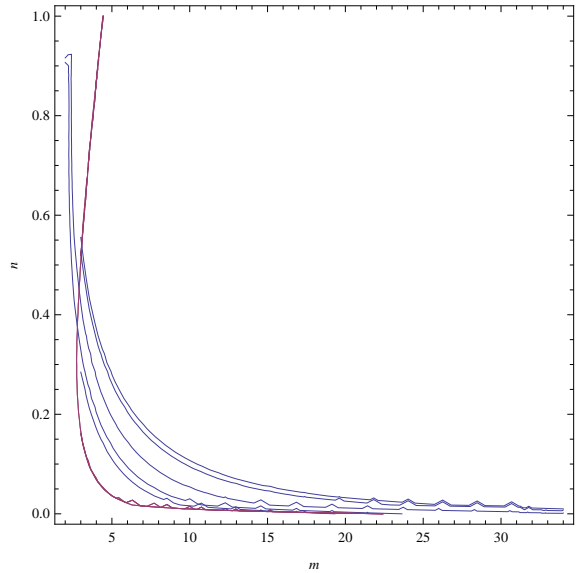}
\caption{The contour plot of $n_s$ and $r$ for the range of values
$n_s=[0.9635,0.968]$ and $r=0.055$, as functions of $m$ and $n$,
with the latter taking values in the ranges $m=[2,34]$ and
$n=[0,1]$, for $(N,\Lambda,\xi,c_1)=(60,2\times
10^{-7},0.14,10^{20})$.} \label{plot1}
\end{figure}

\begin{figure}[h!]
\centering
\includegraphics[width=18pc]{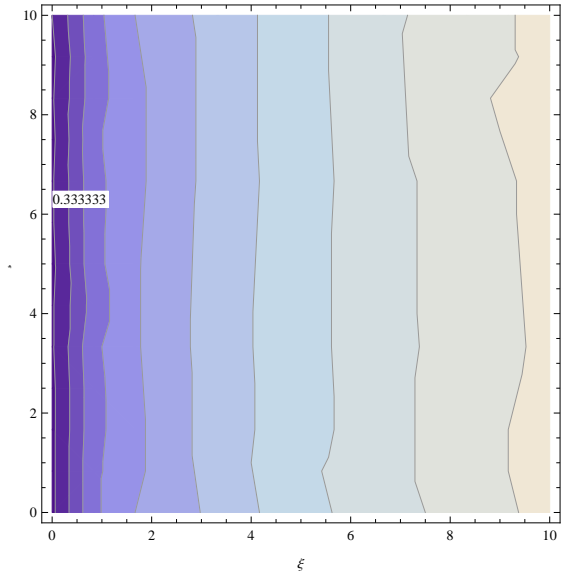}
\caption{The contour plot of the wave speed $c_A^2$ as a function
of $\xi$ and $\Lambda$, for
$(N,c_1,m,n)=(60,10^{20},30.1,10^{-5})$ and for $\Lambda$ and
$\xi$ chosen in the ranges $\Lambda=[0,10]$ and $\xi=[0,10]$.
Darker contours indicate larger values of $c_A^2$.} \label{plot2}
\end{figure}

For simplicity we shall work in reduced Planck units, by setting
$\kappa^2=1$, and a thorough investigation of the parameter space
indicates that the phenomenological viability of the model is
easily achieved. Indeed, a viable phenomenology is obtained for
$(N,\Lambda,\xi,c_1,m,n)=(60,2\times
10^{-7},0.14,10^{20},30.1,10^{-5})$, for which we have,
\begin{equation}\label{viableslowrollcase1}
n_s=0.966682,\,\,\,r=0.0573751\, ,
\end{equation}
and these values are within the Planck 2018 constraints
(\ref{observationaldatanewresults}). Also it should be noted that
for negative values of $c_1$ we obtain complex values for the
observational indices, so only positive values of $c_1$ are
allowed. In addition, $c_1$ must take larger values in reduced
Planck units, since smaller values do not produce a viable
phenomenology, plus the approximations we did do not hold true for
small values of $c_1$. This is the only constraint we found for
the model at hand, since the interplay of the $(\Lambda ,\xi )$
and $(m,n)$ values always produces a viable phenomenology,
provided that $c_1$ takes large values. Another interesting
feature of the model is that for the values of the free parameters
for which the phenomenological viability of the model is ensured,
the slow-roll indices also respect the slow-roll condition, so
these are well below unity, and recall that these are evaluated at
the first horizon crossing. For example if we choose
$(N,\Lambda,\xi,c_1,m,n)=(60,2\times
10^{-7},0.14,10^{20},30.1,10^{-5})$ we get
$|\epsilon_1|=0.00826076$ which is too small, so the slow-roll
assumption indeed holds true during the horizon crossing. The
viability of the model can be obtained for a wide range of the
free parameters. For example in Fig. \ref{plot1} we present the
contour plot of the spectral index and for the tensor-to-scalar
ratio for the range of values $n_s=[0.9635,0.968]$ and $r=0.055$,
as functions of $m$ and $n$, with the latter taking values in the
ranges $m=[2,34]$ and $n=[0,1]$, for
$(N,\Lambda,\xi,c_1)=(60,2\times 10^{-7},0.14,10^{20})$. The
points where the curves meet indicate the values of $(m,n)$ for
which the simultaneous compatibility of the observational indices
with the Planck data (\ref{observationaldatanewresults}) occurs.
As a final task before ending this section, we shall investigate
the values of the wave speed $c_A^2$, for the values of the free
parameters for which the compatibility with the latest Planck data
is obtained. Particularly, if  $c_A^2<0$, then Jeans instabilities
occur in the theory and also superluminal modes may occur if
$c_A^2>1$ \cite{Babichev:2007dw}. A thorough investigation of the
parameter space shows that $0<c_A^2<1$ for a wide range of values,
even for the values of the free parameters for which the model is
not phenomenologically viable. For example, in the case
$(N,\Lambda,\xi,c_1,m,n)=(60,2\times
10^{-7},0.14,10^{20},30.1,10^{-5})$ for which the model is
compatible with the Planck data, the wave speed is equal to
$c_A^2=0.33333$. In order to have a clear picture of the behavior
of the wave speed, in Fig. \ref{plot2} we present the contour plot
of the wave speed $c_A^2$ as a function of $\xi$ and $\Lambda$,
for $(N,c_1,m,n)=(60,10^{20},30.1,10^{-5})$ and for $\Lambda$ and
$\xi$ chosen in the ranges $\Lambda=[0,10]$ and $\xi=[0,10]$. The
darker contours indicate that the values of $c_A^2$ grow larger.
As it can be seen, the wave speed takes in all cases values
$0<c_A^2<1$, so neither Jeans instabilities nor superluminal modes
occur in the theory.
Finally, before closing this section we need to validate that the
assumptions made in Eqs. (\ref{neweqn}) and (\ref{neweqn1}) hold
true, for the values of the free parameters that guarantee the
viability of the model. Indeed by choosing,
$(N,\Lambda,\xi,c_1,m,n)=(60,2\times
10^{-7},0.14,10^{20},30.1,10^{-5})$, in reduced Planck units, the
term $\Big{|}\frac{c_1\dot{\phi}^4}{2}\Big{|}$ reads,
\begin{equation}\label{neweqnanalytic1}
\Big{|}\frac{c_1\dot{\phi}^4}{2}\Big{|}=6.11811\times 10^{-22}\, ,
\end{equation}
and the term $\Big{|}\frac{H
h'(\phi)\dot{\phi}}{\kappa^2}\Big{|}$,
\begin{equation}\label{neweqn1analytic}
\Big{|}\frac{H h'(\phi)\dot{\phi}}{\kappa^2}\Big{|}= 2.0326\times
10^{-28}\, ,
\end{equation}
while the term $\Big{|}\frac{h''(\phi)\dot{\phi}^2}{\kappa^2}
\Big{|}$ is approximately,
\begin{equation}\label{neweqn1analyticnew1}
\Big{|}\frac{h''(\phi)\dot{\phi}^2}{\kappa^2} \Big{|}= 1.1\times
10^{-31}\, ,
\end{equation}
so obviously the assumptions in Eqs. (\ref{neweqn}) and
(\ref{neweqn1}) hold true. In addition, let us also validate that
the assumption of Eq. (\ref{slowrollassumption2}) holds true, so
by also using $(N,\Lambda,\xi,c_1,m,n)=(60,2\times
10^{-7},0.14,10^{20},30.1,10^{-5})$, in reduced Planck units, we
get,
\begin{equation}\label{potentialvalue}
V(\phi)=1.11093\times 10^{-19}
\end{equation}
in reduced Planck units, which is obviously way larger than the
numerical value appearing in Eq. (\ref{neweqnanalytic1}). So we
validated that the approximations we made for a slow-rolling
scalar hold true. We need to note that the slow-roll assumptions
in our case, and in contrast to the standard canonical scalar
field case, invoke the derivatives of the scalar field and the
coupling to the curvature scalar, namely the function $h(\phi)$.

\subsection{Models with Potential: Constant-roll Phenomenology}

Unlike in the slow-roll case, the model of Eq.
(\ref{exponentialmodel}) may become compatible with the latest
Planck data, if the slow-roll assumption for the scalar field,
namely Eq. (\ref{slowrollscalarfield}) is replaced by the
constant-roll assumption,
\begin{equation}\label{constantrollcondition}
\ddot{\phi}=\beta H\dot{\phi}\, ,
\end{equation}
where $\beta$ is a dimensionless free parameter. Thus, in view of
the constant-roll condition (\ref{constantrollcondition}), and
also by still assuming that the condition
(\ref{slowrollassumption2}) holds true, the equations of motion
for the model (\ref{frxfunctionpotentialslowroll}) yield the
following solutions for $\dot{\phi}$, $H$ and $\dot{H}$ as a
function of the scalar field,
\begin{equation}\label{euqationsofmotion11potentialslowroll1const}
H\simeq \frac{\sqrt{\frac{\kappa ^2 V(\phi )}{h(\phi
)}}}{\sqrt{3}}\, ,
\end{equation}
\begin{equation}\label{euqationsofmotion21potentialslowroll2const}
\dot{H}\simeq -\frac{\text{c1} \kappa ^2 \left(\frac{2 V'(\phi
)-\frac{4 V(\phi ) h'(\phi )}{h(\phi )}}{(\beta +1) \text{c1}
\sqrt{\frac{\kappa ^2 V(\phi )}{h(\phi )}}}\right)^{4/3}}{4\
3^{2/3} h(\phi )}\, ,
\end{equation}
\begin{equation}\label{euqationsofmotion31potentialslowroll3const}
\dot{\phi}\simeq -\frac{\sqrt[3]{\frac{2 V'(\phi )-\frac{4 V(\phi
) h'(\phi )}{h(\phi )}}{(\beta +1) \text{c1} \sqrt{\frac{\kappa ^2
V(\phi )}{h(\phi )}}}}}{\sqrt[6]{3}}\, .
\end{equation}
Accordingly, the slow-roll indices (\ref{slowrollparameters}) in
terms of the scalar field and the free parameters of the theory,
including $\beta$ in the case at hand, read,
\begin{align}\label{slowrollparametersconst}
& \epsilon_1\simeq -\frac{\sqrt[3]{3} \text{c1} e^{\kappa  \phi }
\left((2 \gamma +1)^{4/3} \Lambda  e^{\frac{1}{3} 4 \gamma  \kappa
\phi } \left(\kappa ^{4/3} \xi ^{4/6} e^{\frac{1}{6} 4 (-(\gamma
+1)) \kappa  \phi }\right)\right)}{\left(2^{2/3} \xi \right)
\left(\Lambda ^{4/6} \left(\kappa ^{4/3} ((\beta +1)
\text{c1})^{4/3}\right)\right)} \, , \\ \notag &
 \epsilon_2\simeq \beta\,
,\\ \notag & \epsilon_3\simeq \frac{h'(\phi ) \sqrt[3]{\frac{6
V(\phi ) h'(\phi )-3 h(\phi ) V'(\phi )}{(\beta +1) \text{c1}
h(\phi ) \sqrt{\frac{\kappa ^2 V(\phi )}{h(\phi )}}}}}{2^{2/3}
h(\phi ) \sqrt{\frac{\kappa ^2 V(\phi )}{h(\phi )}}} \, ,\\ \notag
& \epsilon_4\simeq \frac{\sqrt[3]{6} (\beta +1) h'(\phi )
\left(\frac{2 V(\phi ) h'(\phi )-h(\phi ) V'(\phi )}{(\beta +1)
\text{c1} h(\phi ) \sqrt{\frac{\kappa ^2 V(\phi )}{h(\phi
)}}}\right)^{2/3} \left(\text{c1} \kappa ^2 \left(\frac{2 V(\phi )
h'(\phi )-h(\phi ) V'(\phi )}{(\beta +1) \text{c1} h(\phi )
\sqrt{\frac{\kappa ^2 V(\phi )}{h(\phi
)}}}\right)^{2/3}+\sqrt[3]{6} h''(\phi )\right)}{\sqrt[3]{6}
(\beta +1) h'(\phi )^2 \sqrt{\frac{\kappa ^2 V(\phi )}{h(\phi )}}
\sqrt[3]{\frac{2 V(\phi ) h'(\phi )-h(\phi ) V'(\phi )}{(\beta +1)
\text{c1} h(\phi ) \sqrt{\frac{\kappa ^2 V(\phi )}{h(\phi )}}}}+4
\kappa ^2 V(\phi ) h'(\phi )-2 \kappa ^2 h(\phi ) V'(\phi )} \, .
\end{align}
Also in this case, the wave speed $c_A$ reads,
\begin{equation}\label{wavespeedconst}
c_A=\sqrt{\frac{\frac{1}{2} 3^{2/3} \gamma ^2 \Lambda  e^{\gamma
\kappa  \phi } \left(\frac{-4 \gamma  \kappa  \xi  e^{-\kappa \phi
}-2 \kappa  \xi  e^{-\kappa  \phi }}{(\beta +1) \text{c1}
\sqrt{\frac{\kappa ^2 \xi  e^{-\gamma  \kappa  \phi -\kappa  \phi
}}{\Lambda }}}\right)^{2/3}+\frac{\text{c1} \left(\frac{-4 \gamma
\kappa  \xi  e^{-\kappa  \phi }-2 \kappa  \xi  e^{-\kappa  \phi
}}{(\beta +1) \text{c1} \sqrt{\frac{\kappa ^2 \xi  e^{-\gamma
\kappa  \phi -\kappa  \phi }}{\Lambda }}}\right)^{4/3}}{2\
3^{2/3}}}{\frac{1}{2} 3^{2/3} \gamma ^2 \Lambda  e^{\gamma  \kappa
\phi } \left(\frac{-4 \gamma  \kappa  \xi  e^{-\kappa  \phi }-2
\kappa  \xi  e^{-\kappa  \phi }}{(\beta +1) \text{c1}
\sqrt{\frac{\kappa ^2 \xi  e^{-\gamma  \kappa  \phi -\kappa  \phi
}}{\Lambda }}}\right)^{2/3}+\frac{1}{2} \sqrt[3]{3} \text{c1}
\left(\frac{-4 \gamma  \kappa  \xi  e^{-\kappa  \phi }-2 \kappa
\xi  e^{-\kappa  \phi }}{(\beta +1) \text{c1} \sqrt{\frac{\kappa
^2 \xi  e^{-\gamma  \kappa  \phi -\kappa  \phi }}{\Lambda
}}}\right)^{4/3}}}\, .
\end{equation}
Following the same procedure as in the previous subsection, we may
easily obtain the spectral index and the tensor-to-scalar ratio as
a function of the $e$-foldings number $N$ and the free parameters
of the theory, including $\beta$, for the model of Eq.
(\ref{exponentialmodel}). However the final expressions are too
lengthy to quote here, so we directly proceed to the analysis of
the phenomenological viability of the model. As in the slow-roll
case, we shall work in reduced Planck units, so by choosing for
example $(N,\Lambda,\xi ,c_1,\gamma ,\beta
)=(60,10^{-20},10^{-6},10^{40},10^{8},0.017)$, we obtain,
\begin{equation}\label{viableslowrollcase1constantroll}
n_s=0.965999,\,\,\,r=3.00219\times 10^{-6}\, ,
\end{equation}
which are compatible with the Planck 2018 constraints
(\ref{observationaldatanewresults}). We need to note that the
viability for the exponential model (\ref{exponentialmodel}) comes
with more difficulty in comparison to the power-law model
(\ref{powerlawmodel}), but this is a model-dependent feature and
has nothing to do with the constant-roll condition. Finally, the
wave speed is also between $0<c_A^2<1$, at least for the values of
the free parameters that guarantee the phenomenological viability
of the model, so no Jeans instabilities or superluminal
propagation of scalar modes occur in this case too. For example,
if we choose $(N,\Lambda,\xi ,c_1,\gamma ,\beta
)=(60,10^{-20},10^{-6},10^{40},10^{8},0.017)$, we have
$c_A^2=0.333333$, which of course satisfies $0<c_A^2<1$.

Having discussed the models with potential, in the next subsection
we proceed to the study of models of the form (\ref{frxfunction})
without scalar potential.

\subsection{Model without Potential: Standard and Constant-roll Evolution}

Let us now consider another model of $k$-inflation, in this case
without scalar potential. The model is of the form of
non-minimally coupled scalar field to gravity, and with only the
higher order kinetic term appearing in the Lagrangian.
Particularly, the $f(R,X,\phi)$ function has the following form,
\begin{equation}\label{frxfunctionmodelinopotential}
f(R,\phi,X)=\frac{h(\phi)R}{\kappa^2}+c_1 X^{2}\, .
\end{equation}
In this subsection we shall consider the phenomenology of the
above model, with respect to its viability, in the constant-roll
case.

We shall assume that the constant-roll condition of Eq.
(\ref{constantrollcondition}) holds true for the scalar field. The
non-constant-roll case can be obtained easily by setting $\beta=0$
in the equations that will follow. In this case, without assuming
any condition on the scalar field, except for the constant-roll
evolution of Eq. (\ref{constantrollcondition}), the expressions
for $H$, $\dot{H}$ and $\dot{\phi}$ as functions of the scalar
field, are,
\begin{equation}\label{euqationsofmotion11potentialslowroll1model1constanrollnopotential}
H= \frac{32 \sqrt{\frac{2}{15}} h'(\phi )^2 \sqrt{\frac{\text{c1}
\kappa ^2}{h(\phi )}}}{3 (\beta +1)^2 \text{c1} \kappa ^2 h(\phi
)}\, ,
\end{equation}
\begin{equation}\label{euqationsofmotion21potentialslowroll2model1constanrollnopotential}
\dot{H}\simeq \Big{(}\frac{H
h'(\phi)\dot{\phi}}{\kappa^2}-\frac{h''(\phi)\dot{\phi}^2}{\kappa^2}-\frac{c_1\dot{\phi}^4}{2}\Big{)}\,
,
\end{equation}
\begin{equation}\label{euqationsofmotion31potentialslowroll3model1constanrollnopotential}
\dot{\phi}= \frac{8 \sqrt{\frac{2}{15}} h'(\phi )}{(\beta +1)
h(\phi ) \sqrt{\frac{\text{c1} \kappa ^2}{h(\phi )}}}\, ,
\end{equation}
and notice that the expressions for $\dot{H}$ given in Eqs.
(\ref{euqationsofmotion21potentialslowroll}) and
(\ref{euqationsofmotion21potentialslowroll2model1constanrollnopotential})
coincide. The slow-roll indices (\ref{slowrollparameters}) as
functions of the scalar field $\phi$ for the model
(\ref{frxfunctionmodelinopotential}), can easily be evaluated by
using equations
(\ref{euqationsofmotion11potentialslowroll1model1constanrollnopotential}),
(\ref{euqationsofmotion21potentialslowroll2model1constanrollnopotential})
and
(\ref{euqationsofmotion31potentialslowroll3model1constanrollnopotential}),
so these read,
\begin{align}\label{slowrollparameters}
& \epsilon_1= -\frac{2 \phi ^{-2 n} \left(\text{c1} n+(\beta +1)
\Lambda ^2 ((4 \beta +5) n-4 (\beta +1)) \phi ^{2
n}\right)}{\Lambda ^2 n}\, , \\ \notag &
 \epsilon_2= \beta\,
,\\ \notag & \epsilon_3=-\frac{2 (\beta +1)^3 \phi ^2 \sqrt{\kappa
^2 \Lambda  \phi ^n} \sqrt{\frac{n^4 \phi ^{-n-4}}{(\beta +1)^4
\kappa ^2 \Lambda }}}{n^2} \, , \\ \notag & \epsilon_4= -\frac{2 n
\left(\text{c1} n+4 (\beta +1)^2 \Lambda ^2 (n-1) \phi ^{2
n}\right)}{(\beta +1) \phi ^2 \sqrt{\kappa ^2 \Lambda  \phi ^n}
\left(\text{c1}+2 (\beta +1)^2 \Lambda ^2 \phi ^{2 n}\right)
\sqrt{\frac{n^4 \phi ^{-n-4}}{(\beta +1)^4 \kappa ^2 \Lambda }}}\,
.
\end{align}
We shall consider the model,
\begin{equation}\label{modelconstantrollonlyviable}
h(\phi)=\Lambda \phi^n\, ,
\end{equation}
which is similar with the case in the presence of a potential, so
we now examine the viability of the present model. Also in this
case, the wave speed $c_A$ of Eq. (\ref{wavespeed}) reads,
\begin{equation}\label{wavespeedmodeleternal}
c_A=\sqrt{\frac{\frac{\text{c1} n^4 \phi ^{-2 n-4}}{8 (\beta +1)^4
\kappa ^4 \Lambda ^2}+\frac{3 n^4}{4 (\beta +1)^2 \kappa ^4 \phi
^4}}{\frac{3 \text{c1} n^4 \phi ^{-2 n-4}}{8 (\beta +1)^4 \kappa
^4 \Lambda ^2}+\frac{3 n^4}{4 (\beta +1)^2 \kappa ^4 \phi ^4}}}\,
,
\end{equation}
Using the $e$-foldings number, which in this case, the term
$\frac{H}{\dot{\phi}}$ in Eq. (\ref{efoldings2}) reads,
\begin{equation}\label{asxeeqn}
\frac{H}{\dot{\phi}}\simeq -\frac{h'(\phi )}{4 (\beta +1) h(\phi
)}\, ,
\end{equation}
so the slow-roll indices, the corresponding observational indices
and the wave speed, shall be evaluated for the following scalar
field value $\phi$ at the horizon crossing,
\begin{equation}\label{horizoncrossingvalue}
\phi_i=2^{\frac{1}{2 n}} \Lambda ^{-1/n} \left(-\frac{\left(8
\beta ^2+18 \beta +9\right) n-8 (\beta +1)^2}{\text{c1}
n}\right)^{-\frac{1}{2 n}} e^{\frac{4 (\beta +1) N}{n}}\, .
\end{equation}
The resulting expression for the observational indices and the
wave speed are quite lengthy to quote here, but a detailed
analysis indicates that neither the constant-roll ($\beta \neq 0$)
not the slow-roll case ($\beta=0$) case, yield any viable results.
In fact, the spectral index can be compatible with the Planck
data, however the tensor-to-scalar ratio is always extremely
higher than the allowed observational data. However, the wave
speed is always nearly equal to unity, for all the values of the
free parameters that yield sensible results (non-imaginary
values). Therefore, we may claim even from now that the power-law
model without scalar potential is not so appealing.

We need to note that in the literature an interesting class of
models is studied \cite{Li:2012vta}, in which case a scalar
potential and only a higher kinetic term is included in the
Lagrangian, without a canonical kinetic term. These theories can
be viable, and in contrast, the non-minimally coupled theory with
only a higher kinetic term is not viable, at least when the
non-minimally coupled function is a power-law type. Perhaps
including higher order kinetic terms, higher than the quadratic
which we studied, may resolve this issue, but we leave this to the
reader, since it is a repetition of the method we provided above,
and the result is highly model dependent, with the dependence
quantified in the choice of the coupling function to the curvature
scalar $h(\phi)$.

\subsection{Conformally Transformed Models with Scalar Potential and Implications}

The viability of the models with scalar potential of the form
(\ref{frxfunctionpotentialslowroll}) has a direct implication for
minimally coupled $k$-inflation models with non-trivial
higher-order kinetic terms, with scalar potential, due to the fact
that these two theories are related via a conformal
transformation. From a phenomenological point of view, the power
spectrum of the primordial scalar curvature perturbations and the
power spectrum of the tensor perturbations are conformal invariant
quantities, so the spectral index of the primordial curvature
perturbations and the tensor-to-scalar ratio for the two theories
are expected to be the same in principle, although deviations from
this rule are presented in the literature, mainly related to the
slow-roll condition
\cite{Odintsov:2018qyy,Karam:2019dlv,Karam:2017zno}. Let us
consider the action of the non-minimally coupled $k$-inflation
theory with scalar potential corresponding to Eq.
(\ref{frxfunctionpotentialslowroll}), which is,
\begin{equation}
\label{mainactionBconformal} \mathcal{S}=\int d^4x\sqrt{-g}\left[
\frac{1}{2}\left( \frac{h(\phi)R}{\kappa^2}-2V(\phi)+c_1\gamma
X^{2}\right) \right]\, ,
\end{equation}
and let us perform a conformal transformation of the metric, which
has the following form,
\begin{equation}\label{conformaltranformation}
g_{\mu \nu}=\Omega^{-2}(x^{\mu})\tilde{g}_{\mu \nu}\, ,
\end{equation}
where $\Omega (x^{\mu})$ is a differentiable function of the
spacetime coordinates, and the ``tilde'' metric denotes the
conformally transformed metric. Accordingly, $\sqrt{-g}$ and
$g^{\mu \nu}$ under the conformal transformation $\Omega
(x^{\mu})$ transform as follows,
\begin{equation}\label{conformaltranformation1}
g^{\mu \nu}=\Omega (x^{\mu})\tilde{g}^{\mu \nu}\, ,
\,\,\,\sqrt{-g}=\Omega^{-4}\sqrt{-\tilde{g}}\, .
\end{equation}
We introduce the notation,
\begin{equation}\label{notation}
f=\ln \Omega (x^{\mu})\, ,
\,\,\,f_{\mu}=\frac{\partial_{\mu}\Omega
}{\Omega}=\partial_{\mu}f\, ,
\end{equation}
and therefore we have, $\tilde{f}^{\mu}=\tilde{g}^{\mu
\nu}f_{\nu}$. Using the conformal transformation rules and the
notation (\ref{notation}), the Ricci scalar under the conformal
transformation, transforms as follows,
\begin{equation}\label{ricciscalarcofnormal}
R=\Omega^2\left(\tilde{R}+6\tilde{\square}f-6\tilde{g}^{\mu
\nu}f_{\mu}f_{\nu}\right)\, ,
\end{equation}
where the d'Alembertian is,
\begin{equation}\label{dalembert}
\tilde{\square}f=\frac{1}{\sqrt{-\tilde{g}}}\partial_{\mu}\left(
\sqrt{-\tilde{g}}\tilde{g}^{\mu \nu}\partial_{\nu}f\right)\, .
\end{equation}
Having the transformation properties at hand, we can conformally
transform each term of the action (\ref{mainactionBconformal})
straightforwardly. The term $\sim h(\phi) R$ transforms as,
\begin{equation}\label{firstterm}
\frac{1}{2}\sqrt{-g}h(\phi)R\to
\frac{1}{2}\sqrt{-\tilde{g}}h(\phi)\Omega^{-2}\left(
\tilde{R}+6\tilde{\square}f-6\tilde{g}^{\mu
\nu}f_{\mu}f_{\nu}\right)\, .
\end{equation}
Without loss of generality, since $\Omega$ is arbitrary, we can
choose,
\begin{equation}\label{specialcondition}
h(\phi)\Omega^{-2}=1\, ,
\end{equation}
so it is obvious that with this choice, the conformally
transformed frame becomes the Einstein frame due to the occurrence
of the term $\sim \sqrt{-\tilde{g}}\tilde{R}$. The second term in
the Einstein frame which is $\sim \tilde{\square}f$ is a total
derivative, so it disappears by integrating, subject to the
constraint that the integral of the function $f$ on the boundary
of the spacetime vanishes. For the choice
(\ref{specialcondition}), we have,
\begin{equation}\label{specialcaseofderivativef}
f_{\mu}=\frac{1}{2}\frac{\partial_{\mu}h}{h}=\frac{h'}{2h}\partial_{\mu}\phi\,
,
\end{equation}
where the ``prime'' indicates differentiation with respect to the
scalar field. Hence, with the choice (\ref{specialcondition}) in
conjunction with (\ref{specialcaseofderivativef}), the
transformation for the first term finally becomes,
\begin{equation}\label{firsttermnewnew}
\frac{1}{2}\sqrt{-g}h(\phi)R\to \frac{1}{2}\sqrt{-\tilde{g}}\left(
\tilde{R}-\frac{3}{2}\left(\frac{h'}{h}\right)^2\tilde{g}^{\mu
\nu}\partial_{\mu}\phi \partial_{\nu}\phi \right)\, .
\end{equation}
Accordingly, the potential term $\sim -2V(\phi)$ transforms as,
\begin{equation}\label{potentialtermtransformation}
\frac{-\sqrt{-g}}{2}\left(-2V(\phi)\right) \to
\frac{-\sqrt{-\tilde{g}}}{2}\frac{1}{h^2}\left(-2V(\phi) \right)\,
,
\end{equation}
and the higher order kinetic term $\sim c_1 X^2$ transforms as,
\begin{align}\label{hgiherorderkineticterm}
&\frac{\sqrt{-g}}{2}c_1X^2=\frac{\sqrt{-g}}{2}c_1\left(\frac{1}{2}g^{\mu
\nu}\partial_{\mu}\phi\partial_{\nu}\phi\right)^2\\
\notag &
=\frac{\sqrt{-\tilde{g}}}{2}\Omega^{-4}c_1\left(\frac{1}{2}\Omega^2\tilde{g}^{\mu
\nu}\partial_{\mu}\phi\partial_{\nu}\phi\right)^2=\frac{\sqrt{-\tilde{g}}}{2}c_1\left(\frac{1}{2}\tilde{g}^{\mu
\nu}\partial_{\mu}\phi\partial_{\nu}\phi\right)^2\\ \notag &
=\frac{\sqrt{-\tilde{g}}}{2}c_1X^2\, ,
\end{align}
so it is basically the same as in the Jordan frame. This result is
an artifact of the choice $\mu=2$ in Eq. (\ref{frxfunction}), and
for a general $\mu$, the conformal transformation would read,
\begin{align}\label{hgiherorderkinetictermconfgeneral}
&\frac{\sqrt{-g}}{2}c_1X^{\mu}=\frac{\sqrt{-g}}{2}c_1\left(\frac{1}{2}g^{\mu
\nu}\partial_{\mu}\phi\partial_{\nu}\phi\right)^{\mu}\\
\notag &
=\frac{\sqrt{-\tilde{g}}}{2}\Omega^{-4}c_1\left(\frac{1}{2}\Omega^2\tilde{g}^{\mu
\nu}\partial_{\mu}\phi\partial_{\nu}\phi\right)^{\mu}=\frac{\sqrt{-\tilde{g}}}{2}c_1\left(\frac{1}{2}\tilde{g}^{\mu
\nu}\partial_{\mu}\phi\partial_{\nu}\phi\right)^{\mu}\\ \notag &
=\frac{\sqrt{-\tilde{g}}}{2}\Omega^{2(-2+\mu )}c_1X^{\mu}\, .
\end{align}
Therefore, the conformally transformed action
(\ref{mainactionBconformal}) in the Einstein frame for general
$\mu$ reads,
\begin{equation}
\label{mainactionBconformallytranformed} \mathcal{S}=\int
d^4x\sqrt{-\tilde{g}}\frac{1}{2}\left[
\tilde{R}-\frac{3}{2}\left(\frac{h'}{h}\right)^2\tilde{g}^{\mu
\nu}\partial_{\mu}\phi \partial_{\nu}\phi -\frac{2}{h^2}V(\phi)
+h(\phi )^{(-2+\mu )}c_1X^{\mu} \right]\, .
\end{equation}
The above action can easily be transformed to an Einstein frame
scalar $k$-inflation theory with canonical kinetic term, by
appropriately redefining the scalar field. Thus, due to the
conformal relation between the actions
(\ref{mainactionBconformal}) and
(\ref{mainactionBconformallytranformed}), and owing to the fact
that both the  scalar and tensor power spectrum is conformal
invariant, the two theories can produce similar phenomenology.
This fact indicates that the model
(\ref{mainactionBconformallytranformed}) with $\mu=2$ can be a
viable model, and we can conclude this without performing the
actual calculations, which would be more tedious in comparison to
the model (\ref{mainactionBconformal}). Similar results can hold
true for general $\mu$ but we do not present here for brevity.

\section{Conclusions}

In this paper we studied in depth several $k$-inflation models
with non-minimal coupling of the scalar field to the scalar
curvature. We mainly focused on models that lack of a canonical
kinetic term for the scalar field, in the presence and the absence
of a scalar potential, and we studied two cases with regard to the
evolution of the scalar field, the slow-roll and the constant-roll
case. In the presence of a scalar potential, and with the higher
order kinetic term being of the form $\sim c_1 X^2$, when the
slow-roll conditions are assumed for the scalar field, we
demonstrated that a phenomenologically viable theory can be
obtained. Particularly, we calculated the slow-roll indices and
the observational indices and we showed that the results can be
compatible with the latest Planck constraints. This result is of
course model dependent, but it is quite general since it can hold
true for several choices of the potential and the non-minimal
coupling function $h(\phi)$. Also the sound speed of the model
never exceeds unity for a wide range of the free parameters of the
model, therefore no superluminal propagation occurs, and also it
is positive and hence no instabilities occur. Similar results are
obtained when the constant-roll condition is assumed for the
model. In the case that the scalar potential is absent, the
power-law model is not viable, however the results are model
dependent, and perhaps the inclusion of higher order kinetic terms
may render the model phenomenologically viable. Finally, we
performed a conformal transformation of the non-minimally coupled
$k$-inflation theory with potential, and the resulting Einstein
frame theory contains a scalar potential, the higher-order kinetic
term and an ordinary scalar kinetic term. Due to the conformal
invariance of both the scalar and tensor power spectra, we argued
that the resulting Einstein frame can also be viable and
compatible with the observational data. The latter theory is quite
more involved to handle analytically, thus by studying the
non-minimally coupled model, one may obtain several
phenomenological conclusions for the Einstein frame theory.
Nevertheless, the most correct way to prove the equivalence is to
calculate the observational indices, using the slow-roll or the
constant-roll condition, due to the fact that the very slow-roll
condition may obscure the equivalence of the two theories, see for
example \cite{Odintsov:2018qyy,Karam:2019dlv,Karam:2017zno}. This
task is not easy in general though, but we hope to address it in a
forthcoming work.

\end{document}